\input{aipcheck}

\documentclass[
,final            
  ]
  {aipproc}

\layoutstyle{8x11double}

\begin{document}

\title{$g$-factor anisotropy in nanowire-based InAs quantum dots }


\author{Samuel d'Hollosy}{
  address={Departement of Physics, University of Basel, Klingelbergstr. 82, CH-4056 Basel, Switzerland}
}

\author{G\'{a}bor F\'{a}bi\'{a}n}{
  address={Departement of Physics, University of Basel, Klingelbergstr. 82, CH-4056 Basel, Switzerland}
}

\author{Andreas Baumgartner}{
  address={Departement of Physics, University of Basel, Klingelbergstr. 82, CH-4056 Basel, Switzerland}
}

\author{Jesper Nyg\r{a}rd}{
  address={Niels Bohr Institute, University of Copenhagen, Universitetsparken 5, DK-2100 Copenhagen, Denmark }
}

\author{Christian Sch\"{o}nenberger}{
  address={Departement of Physics, University of Basel, Klingelbergstr. 82, CH-4056 Basel, Switzerland}
}

\begin{abstract}
The determination and control of the electron $g$-factor in semiconductor quantum dots (QDs) are fundamental prerequisites in modern concepts of spintronics and spin-based quantum computation. We study the dependence of the $g$-factor on the orientation of an external magnetic field in quantum dots (QDs) formed between two metallic contacts on stacking fault free InAs nanowires. We extract the $g$-factor from the splitting of Kondo resonances and find that it varies continuously in the range between $|g^*| = 5$ and 15.
\end{abstract}

\maketitle

InAs nanowires (NWs) are a versatile material basis for a large variety of fundamental phenomena, e.g. Cooper pair splitting \cite{Hofstetter2009, Hofstetter2011}. The large electron $g$-factor inherited from bulk InAs leads to energetically separated spin states already in relatively small magnetic fields. The spins of the confined electrons in NW quantum dots (QDs) can exhibit significant variations of the $g$-factor from bulk to near vacuum values \cite{Pryor2006}. A tunable $g$-factor is crucial for studying the spin dynamics in nanostructures, or for the implementation of quantum computation on QDs \cite{Loss1998}. 
Here we present transport spectroscopy measurements on nominally stacking fault free InAs NWs. We extract the $g$-factor anisotropy by analyzing the splitting of a Kondo resonance as a function of the orientation of an external magnetic field and find variations of up to a factor of 3. Similar experiments using the Zeeman splitting of Coulomb resonances can be found in Ref.~\cite{Deacon2011}. In addition, we show that the Kondo resonance splitting can exhibit different characteristics, depending on the angle of the external field.

\begin{figure}[t]
  \includegraphics[width=\columnwidth]{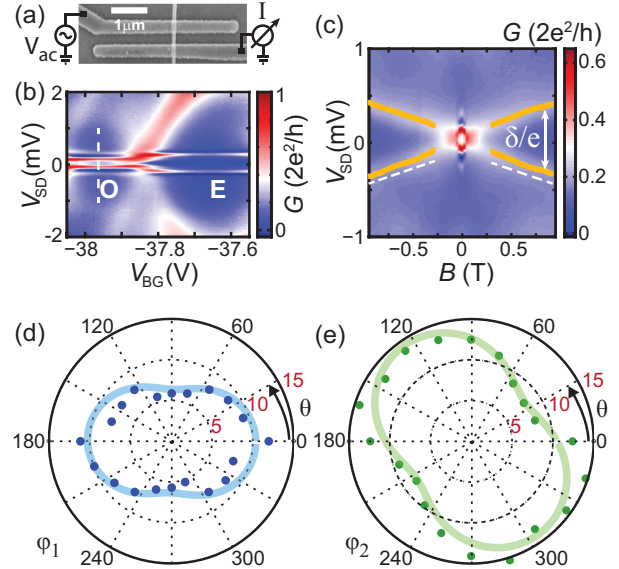}
  \caption{(a) SEM image of a device and simplified schematic of the measurement setup. (b) Differential conductance $G$ as a function of the backgate ($V_{\rm BG}$) and the bias voltage ($V_{\rm SD}$). The dashed line indicates the gate voltage of the measurements in figure (c). (c) $G$ vs. $V_{\rm SD}$ and $B$ at angles $\varphi_1=-25^\circ$ and $\theta=72^\circ$. Yellow crosses indicate the peak positions of the split Kondo resonance. (d) and (e) $g$-factor extracted from the Kondo splitting as a function of the angle $\theta$ at the settings $\varphi_1=-25^\circ$ and $\varphi_2=+20^\circ$, respectively. The continuous curves
 stem from the fit explained in the text.}
\end{figure}

A typical device is shown in Fig.~1a: an InAs nanowire is connected to two Ti($5\,$nm)/Al($100\,$nm) superconducting contacts separated by $\sim250\,$nm. Devices are fabricated on top of a $400\,$nm $\mathrm{SiO_2}$ layer. A voltage $V_{\rm BG}$ is applied to the highly doped Si substrate, which serves as backgate.
We measure the differential conductance $G$ with standard lock-in techniques and superimpose a dc voltage $V_{\rm SD}$ to the ac excitation $V_{\rm ac}$ for spectroscopy measurements. A vector magnet allows us to apply an external magnetic field $B$ in any direction relative to the NW axis. The experiments were performed in a dilution refrigerator at $\sim60\,$mK base temperature.

In Fig~1b, $G$ is plotted as a function of $V_{\rm BG}$ and $V_{\rm SD}$. Experiments at $B=200\,$mT are suppressing the superconductivity (not shown) and allow us to identify a Kondo resonance in the odd charge state labeled "O".
The corresponding features in the superconducting state in Fig.~1b are consistent with Kondo modulated Andreev transport discussed in Ref.~\cite{Jespersen2007}. Here we focus on the splitting $\delta$ of the Kondo resonance in an external magnetic field. In Fig.~1c, $G$ is plotted as a function of $B$ and $V_{\rm SD}$ for $V_{\rm BG}=-37.95\,$V. At low fields the contacts are superconducting and the field evolution is given by the critical field of the superconductor. At $B>150\,$mT the superconductivity is suppressed and a clear Kondo resonance develops.
We use the position of the peak maximum to measure the energy splitting $\delta$, which is expected to scale linearly with $B$. This allows one to extract the effective $g$-factor $g^*$ using $\delta = 2|g^*| \mu_B B$ \cite{Meir1993}. Linear fits were done for $250\,$mT$<B<500\,$mT. We repeat this procedure for different angles $\theta$ of the magnetic field in two planes given by the angles $\varphi_1=-25^\circ$ and $\varphi_2=20^\circ$ with respect to the NW axis. The resulting $g^*$ values in the two planes are plotted  in Figs.~1d and 1e.

The $g$-factor as a function of the field orientation is well described by a second order tensor
\[
g(\textbf{B})=\frac{1}{|\textbf{B}|}\sqrt{g_1^{\,2}B_1^{\,2}+g_2^{\,2}B_2^{\,2}+g_3^{\,2}B_3^{\,2}}
\]
with the principle axes rotated with respect to the NW orientation. 
Therefore the fit parameters are the principle $g$-factors $g_i$ and the Euler angles of the rotation. Good agreement is obtained for $g_i$ between 5 and 15. In contrast to previous work \cite{Csonka2008,Schroer2011}, we do not observe $g$-factors larger than than the bulk value ($|g^*|=14.7$). The reduction of the $g$-factor can be explained by quenching of the orbital angular momentum due to quantum confinement \cite{Pryor2006}. For strong confinements (small QDs) the $g$-factor can even approach the free electron value $+2$ \cite{Pryor2006}. We do not observe a clear correlation between the principal axes of the $g$-tensor and the NW orientation, possibly because the confinement potential along the NW ($<250\,$nm) is not much different to the perpendicular directions ($\sim100\,$nm) and thus might be dominated by mesoscopic details.

\begin{figure}[b]
  \includegraphics[width=\columnwidth]{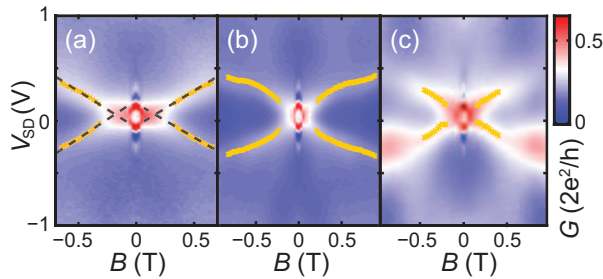}
  \caption{$G$ as function of $V_{SD}$ and $B$ showing a qualitatively different evolution of the Kondo resonance. (a) Splitting linear in $B$ with an offset of $\Delta B\approx 150\,$mT ($\varphi_1$, $\theta=0^\circ$). (b) Non-linear splitting in $B$ ($\varphi_1$, $\theta=-72^\circ$). (c) strongly varying peak conductances, asymmetric in $V_{\rm SD}$ ($\varphi_2$, $\theta=72^\circ$).}
\end{figure}

In our experiments we find rather different field evolution characteristics of the Kondo splitting for different field orientations. Figure~2 shows three additional examples. For ($\varphi_1$, $\theta=0$, $g^*\approx 12$) in Fig.~2a the splitting is linear in $B$, but with an offset of $\Delta B\approx 150\,$mT,
while for ($\varphi_1$, $\theta=-72^\circ$, $g^*\approx 5.5$) in Fig.~1b the evolution is not linear in $B$. For ($\varphi_2$, $\theta=72^\circ$, $g^*\approx 11$) in Fig.~1c the amplitudes of the two split Kondo resonances do not evolve monotonously with $B$ and are not symmetric with respect to $V_{\rm SD}=0$.

These latter observations we tentatively attribute to a modification of the Kondo state by another transport process becoming available at finite bias. One possibility is that inelastic processes start contributing to the formation of the spin-dependent resonance, which might also account for the bias-asymmetry in asymmetrically biased samples. We have no explanation for the low-field offset of the Kondo splitting.

In summary, we report maps of the angle dependence of the $g$-factor in stacking fault free InAs nanowire quantum dots. We use the splitting of a Kondo resonance in a magnetic field to extract the $g$-factor. It varies continuously with the orientation of the external magnetic field, well described by a second order tensor with principle values between $5$ and $15$. In addition, we document different splitting characteristic of the Kondo resonance in the external field.

\begin{theacknowledgments}
This work was financially supported by the EU FP7 project SE2ND, the EU ERC project QUEST, the Swiss NCCR Nano and NCCR Quantum and the Swiss SNF.
\end{theacknowledgments}

\bibliographystyle{aipproc}   
\bibliography{icps_proceeding}

\end{document}